\documentclass[aps,preprint,superscriptaddress,longbibliography]{revtex4-1}%
\usepackage{amsmath,amssymb}
\usepackage{graphicx}
\usepackage{epstopdf}
\usepackage{bm}
\usepackage{color}
\usepackage{amsmath}
\usepackage{amsfonts}
\usepackage{amssymb}%
\providecommand{\U}[1]{\protect\rule{.1in}{.1in}}

\begin{document}

\title{Tailoring high-$T_N$ interlayer antiferromagnetism in a van der Waals itinerant magnet}

\author{Junho Seo}
\thanks{equal contribution}
\affiliation{Center for Artificial Low Dimensional Electronic Systems, Institute for Basic Science (IBS), Pohang 37673, Korea}
\affiliation{Department of Physics, Pohang University of Science and Technology (POSTECH), Pohang 37673, Korea}
\author{Eun Su An}
\thanks{equal contribution}
\affiliation{Center for Artificial Low Dimensional Electronic Systems, Institute for Basic Science (IBS), Pohang 37673, Korea}
\affiliation{Department of Physics, Pohang University of Science and Technology (POSTECH), Pohang 37673, Korea}
\author{Taesu Park}
\thanks{equal contribution}
\affiliation{Department of Chemistry, Pohang University of Science and Technology (POSTECH), Pohang 37673, Korea}
\author{Soo-Yoon Hwang}
\affiliation{Department of Materials Science and Engineering, Pohang University of Science and Technology (POSTECH), Pohang 37673, Korea}
\author{Gi-Yeop Kim}
\affiliation{Department of Materials Science and Engineering, Pohang University of Science and Technology (POSTECH), Pohang 37673, Korea}
\author{Kyung Song}
\affiliation{Materials Modeling and Characterization Department, KIMS, Changwon 51508, Korea}
\author{Eunseok Oh}
\affiliation{Center for Artificial Low Dimensional Electronic Systems, Institute for Basic Science (IBS), Pohang 37673, Korea}
\affiliation{Department of Physics, Pohang University of Science and Technology (POSTECH), Pohang 37673, Korea}
\author{Minhyuk Choi}
\affiliation{Center for Artificial Low Dimensional Electronic Systems, Institute for Basic Science (IBS), Pohang 37673, Korea}
\affiliation{Department of Physics, Pohang University of Science and Technology (POSTECH), Pohang 37673, Korea}
\author{Kenji Watanabe}
\affiliation{Research Center for Functional Materials,
National Institute for Materials Science, 1-1 Namiki, Tsukuba 305-0044, Japan}
\author{Takashi Taniguchi}
\affiliation{International Center for Materials Nanoarchitectonics,
National Institute for Materials Science, 1-1 Namiki, Tsukuba 305-0044, Japan}
\author{Youn Jung Jo}
\affiliation{Department of Physics, Kyungpook National University, Daegu  41566, Korea}
\author{Han Woong Yeom}
\affiliation{Center for Artificial Low Dimensional Electronic Systems, Institute for Basic Science (IBS), Pohang 37673, Korea}
\affiliation{Department of Physics, Pohang University of Science and Technology (POSTECH), Pohang 37673, Korea}
\author{Si-Young Choi}
\email{youngchoi@postech.ac.kr}
\affiliation{Department of Materials Science and Engineering, Pohang University of Science and Technology (POSTECH), Pohang 37673, Korea}
\author{Ji Hoon Shim}
\email{jhshim@postech.ac.kr}
\affiliation{Department of Physics, Pohang University of Science and Technology (POSTECH), Pohang 37673, Korea}
\affiliation{Department of Chemistry, Pohang University of Science and Technology (POSTECH), Pohang 37673, Korea}
\author{Jun Sung Kim}
\email{js.kim@postech.ac.kr}
\affiliation{Center for Artificial Low Dimensional Electronic Systems, Institute for Basic Science (IBS), Pohang 37673, Korea}
\affiliation{Department of Physics, Pohang University of Science and Technology (POSTECH), Pohang 37673, Korea}
\date{\today}

\begin{abstract}
\textbf{
Antiferromagnetic (AFM) van der Waals (vdW) materials~\cite{Burch2018, Gibertini2019, Gong2019, Susner2017, Huang2017, Huang2018, Jiang2018, Song2018, Wang2019, Chen2019, Li2019, Song2019, Sivadas2018, Jiang2019, Jang2019, McGuire2017, Lei2020, Otrokov2019, Deng2020, Wu2019} provide a novel platform for synthetic AFM spintronics~\cite{Jungwirth2016, Baltz2018, Zelezny2018, Duine2018}, in which the spin-related functionalities are derived from manipulating spin configurations between the layers.
Metallic vdW antiferromagnets~\cite{Lei2020, Deng2020, Otrokov2019, Wu2019} are expected to have several advantages over the widely-studied insulating counterparts~\cite{Susner2017, Huang2017, Huang2018, Jiang2018, Song2018, Wang2019, Chen2019, Li2019, Song2019, Sivadas2018, Jiang2019, Jang2019, McGuire2017} in switching and detecting the spin states through electrical currents~\cite{Wang2019_2, Marti2014, Bodnar2018, Wadley2016, Galceran2016, Kriegner2016}, but have been much less explored due to the lack of suitable materials. Here, utilizing the extreme sensitivity of the vdW interlayer magnetism to material composition, we report the itinerant antiferromagnetism in Co-doped Fe$_4$GeTe$_2$ with $T_N$ $\sim$ 210 K, an order of magnitude increased as compared to other known AFM vdW metals~\cite{Lei2020, Deng2020, Otrokov2019, Wu2019}. The resulting spin configurations and orientations are sensitively controlled by doping, magnetic field, temperature, and thickness, which are effectively read out by electrical conduction. These findings manifest strong merits of metallic vdW magnets with tunable interlayer exchange interaction and magnetic anisotropy, suitable for AFM spintronic applications.
}
\end{abstract}
\maketitle
Antiferromagnets have gained more attraction recently as active elements for spintronic applications owing to their advantageous properties over ferromagnets, including negligible stray field, robustness against magnetic perturbation, and ultrafast spin dynamics~\cite{Jungwirth2016, Baltz2018, Zelezny2018, Duine2018}.
One versatile systems for antiferromagnetic (AFM) spintronics is the so-called synthetic antiferromagnets, where ferromagnetic (FM) and non-magnetic layers are stacked alternately using the thin-film deposition techniques~\cite{Duine2018}.
In these magnetic multilayers, the AFM exchange coupling across the FM layers is essentially determined by Ruderman-Kittel-Kasuya-Yoshida (RKKY) interaction, and thus can be more easily modulated than in crystalline antiferromagnets~\cite{Duine2018}. However this also means that interface roughness or chemical intermixing, inevitable and often uncontrollable during the preparation process, governs the resulting spin configurations and properties of synthetic antiferromagnets.
In this regard, van der Waals (vdW) antiferromagnets~\cite{Burch2018, Gibertini2019, Gong2019, Susner2017, Huang2017, Huang2018, Jiang2018, Song2018, Wang2019, Chen2019, Li2019, Song2019, Sivadas2018, Jiang2019, Jang2019, McGuire2017, Lei2020, Otrokov2019, Deng2020, Wu2019} offer an intrinsic magnetic multilayer system with highly-crystalline and atomically-flat interface. They are known to host various types of magnetic ground states~\cite{Burch2018, Gibertini2019, Gong2019, Susner2017, Huang2017, Huang2018, Jiang2018, Song2018, Wang2019, Chen2019, Li2019, Song2019, Sivadas2018, Jiang2019, Jang2019, McGuire2017, Lei2020, Otrokov2019, Deng2020, Wu2019} and can be easily assembled with other vdW materials, which may lead to AFM spintronics based on vdW materials and their heterostructures.

One key issue for vdW-material-based AFM spintronics is to identify the suitable candidate materials. Most of the known vdW antiferromagnets are insulating~\cite{Susner2017, Huang2017, Huang2018, Jiang2018, Song2018, Wang2019, Chen2019, Li2019, Song2019, Sivadas2018, Jiang2019, Jang2019, McGuire2017}, and their interlayer magnetic coupling is mainly through superexchange-like interaction~\cite{Sivadas2018, Jiang2019, Jang2019}. Thus the modulation of interlayer coupling usually requires structural modification~\cite{Sivadas2018, Jiang2019, Jang2019, Li2019, Song2019, Chen2019} and is relatively difficult as compared to synthetic antiferromagnets with the interlayer RKKY coupling~\cite{Duine2018}.
Obviously metallic vdW antiferromagnets~\cite{Lei2020, Deng2020, Otrokov2019, Wu2019} can be a good alternative. The conduction electrons mediate the interlayer interaction, similar to synthetic antiferromagnets, which is expected to be strongly modulated by changing the composition or the interlayer distance. Furthermore their longitudinal or transverse conductivities are sensitive to the spin configurations~\cite{Wang2019_2, Marti2014, Bodnar2018, Wadley2016, Galceran2016, Kriegner2016}, offering a direct probe to the spin states even for a-few-nanometer-thick crystals. Despite these merits, metallic vdW antiferromagnets are rare in nature, except a few recent examples of GdTe$_3$ and MnBi$_2$Te$_4$ showing a low Neel temperature $T_N$ below 30 K~\cite{Lei2020, Deng2020, Otrokov2019, Wu2019}.
Here we show that an iron-based vdW material, Fe$_4$GeTe$_2$ with Co doping, hosts the interlayer AFM phase with a far higher $T_N$ = 210 K. Its spin configuration is found to be effectively controlled and read out by conduction electrons, endowing Co-doped Fe$_4$GeTe$_2$ with a promising role in vdW-material-based spintronics.

We consider iron-based metallic vdW ferromagnets Fe$_n$GeTe$_2$ ($n$ = 3-5) as a vdW analog of the synthetic multilayer systems (Fig. 1a)~\cite{Deiseroth2006, Chen2013, Deng2018, Kim2018, Fei2018, Stahl2018, May2019, Seo2020}. The first known member, Fe$_3$GeTe$_2$ is experimentally identified as a ferromanget with $T_c$ = 220 K~\cite{Deiseroth2006, Chen2013, Deng2018, Kim2018, Fei2018}, but is theoretically predicted to be interlayer-AFM via RKKY-like interaction~\cite{Jang2019_2}. The interlayer AFM coupling is however extremely fragile to the small amount of defects or dopants~\cite{Jang2019_2}, and the AFM phase is mostly inaccessible in real compounds. An alternative candidate is Fe$_4$GeTe$_2$, recently identified as a high $T_c$ metallic ferromagnet with $T_c$ = 270 K ~\cite{Seo2020}. Because of the relatively thick slabs containing two Fe-Fe dumbbells (Fig. 1b), it has stronger intralayer FM coupling than Fe$_3$GeTe$_2$, together with the interlayer FM coupling, confirmed theoretically and experimentally~\cite{Seo2020}. By replacing 1/3 of Fe atoms with Co atoms, however, we found that it becomes antiferromagnetic, as evidenced by its temperature dependent susceptibility $\chi(T)$ (Fig. 1c). Hereafter (Fe,Co)$_4$GeTe$_2$ denotes the AFM compound (Fe$_{1-x}$Co$_{x}$)$_4$GeTe$_2$ ($x$ $=$ 0.33) unless the doping level $x$ is otherwise specified. A clear cusp in $\chi(T)$ taken under a magnetic field $H = 1$ kOe along the $c$ axis reveals AFM transition at the Neel temperature $T_N$ = 210 K. The temperature dependent resistivity $\rho(T)$ is metallic with a slight upturn at low temperatures (Fig. 1c). The conductivity of (Fe,Co)$_4$GeTe$_2$ is $\sim$ $3\times 10^5$ ohm$^{-1}$m$^{-1}$, comparable with that of the pristine Fe$_4$GeTe$_2$~\cite{Seo2020}. These characteristics make (Fe,Co)$_4$GeTe$_2$ a unique vdW AFM metal with the highest $T_N$ among vdW antiferromagnets (Supplementary Table. S1).

The crystal structure of (Fe,Co)$_4$GeTe$_2$ is the same as Fe$_4$GeTe$_2$ in a rhombohedral structure (space group $R\overline{3}m$).
Scanning transmission electron microscopy (STEM) image of (Fe,Co)$_4$GeTe$_2$ crystal visualizes the structural units of Fe-Fe dumbbells alternately above and below the plane of Ge atoms. These Fe-Fe-Ge-Fe-Fe layers are encapsulated with Te atoms,
which shows the similar structural tendency to the pristine Fe$_4$GeTe$_2$ (the inset in Fig. 1d).
A clear vdW gap between the layers is observed without any signature of stacking change or intercalated atoms throughout a wide region (Fig. 1d).
These results imply that Co atoms are dominantly substituted to the Fe sites, not in a type of the interstitial sites.
In Fig. 1e, electron energy loss spectroscopy (EELS) analysis visualizes the chemical information within a monolayer, which represents that Co atoms are homogeneously doped in all the Fe sites.
Co doping results in the reduction of both the in-plane ($a = 4.08 \,\pm \,0.04 \,\rm \AA$) and out-of-plane lattice parameters ($c = 29.10 \,\pm \,0.28 \,\rm \AA$), which are deduced from the selected-area-diffraction-pattern (SADP) analysis, in good agreement with X-ray diffraction results (Supplementary Fig. S1).

Having established the high-$T_N$ AFM phase in (Fe,Co)$_4$GeTe$_2$, we focus the systematic changes of
the magnetic and electrical properties of (Fe$_{1-x}$Co$_x$)$_4$GeTe$_2$ single crystals with a variation of Co doping (0 $\leq$ $x$ $\leq$ 0.39). For $x$ = 0, the FM transition with in-plane alignment of magnetic moments occurs at $T_c$ = 270 K, which is followed by the spin-reorientation transition to the out-of-plane alignment at $T_{\rm SR}$ = 110 K~\cite{Seo2020}. Co doping quickly suppresses the spin-reorientation transition, while keeping the FM transition almost intact with a nearly constant $T_c$ for $x$ $\leq$ 0.23. Upon further Co doping, however, the AFM order develops from $T_N$ = 155 K, well below $T_c$ for $x$ = 0.26 and eventually becomes dominant with high $T_N$ up to 226 K for $x$ = 0.39. The saturation magnetization $M_{\rm sat}$ monotonically decreases with Co doping from 7.1 $\mu_B$/f.u. ($x$ = 0) to 5.5 $\mu_B$/f.u. ($x$ = 0.39) due to magnetic dilution effect (Figs. 2a and 2c).
Concomitantly, the out-of-plane saturation field $H^c_{\rm sat}$ gradually increases with Co doping in the FM phases for $x$ $<$ 0.26. This is consistent with the changes of the magnetic anisotropic energy ($K$) from the easy-axis to the easy-plane types, following $H^c_{\rm sat}$ = 2$K$/$M_{\rm sat}$ (Fig. 2d). Entering the AFM phase, however, $H^c_{\rm sat}$ is determined by the AFM coupling $J$ as described by $H^c_{\rm sat}$ $\approx$ 2$J$/$M_{\rm sat}$ and thus suddenly enhanced up to $\sim$ 6 T for $x$ = 0.39. In the AFM phase, the spin-flop transition is observed for $H \parallel ab$ at $x$ = 0.33, but $H \parallel c$ at $x$ = 0.39 (Supplementary Fig. S4), indicating the easy-plane and the easy-axis type spin alignments, respectively. For $x$ = 0.39, we found another magnetic transition occurs to the unknown phase below $T$ = 90 K (Supplementary Fig. S2).
The resulting phase diagram is summarized in Fig. 2g, which manifests that the magnetic configuration and also the magnetic anisotropy of (Fe$_{1-x}$Co$_x$)$_4$GeTe$_2$ are highly sensitive to Co doping.

The first principles calculations consistently predict the FM-to-AFM phase transition with Co doping. Total energy calculations for the FM and various AFM phases confirm that the FM phase is initially stable at low $x$, but eventually becomes unstable against the AFM phase at high $x$. The most stable AFM phase is the so-called A-type, in which all the spin moments in the whole slab of (Fe,Co)$_4$GeTe$_2$ are ferromagnetically aligned, but across the vdW gap they are antiferromagnetically coupled. This is consistent with the positive Curie-Weiss temperature from the inverse susceptibility in the AFM phase, suggesting the dominant FM interaction within the layers (Supplemenatary Fig. S2). Figure 2f shows the total energy difference $\Delta E$ = $E_{\rm FM} - E_{\rm AFM}$ as a function of Co doping $x$, assuming random distribution of Co dopants over all Fe sites. The FM-to-AFM transition occurs at the critical doping level $x_c$ $\sim$ 0.2 (Fig. 2f). The estimated $x_c$ as well as $M_{\rm sat}$ from calculations (Fig. 2c) is in reasonable agreement with experiments. This indicates that the evolution of the magnetic phase with Co doping is well captured by first principles calculations.

Detailed band structure calculations reveal that Co doping affects significantly the density-of-states (DOS) near the Fermi level. In the nonmagnetic calculations, we found a strong DOS peak in the vicinity of the Fermi level (Supplementary Fig. S5). The resulting strong Stoner instability favors the ferromagnetism within the layer, and  the ferrmagnetically-aligned moment at each layer can be treated as a single localized Heisenberg spin, coupled through the interlayer coupling. This interlayer coupling is however determined by subtle balance of pair exchange interactions between Fe/Co atoms across the vdW gap, which is sensitive to the details in the states at the Fermi level. Using total energy calculations on various interlayer spin configurations (Supplementary Fig. S6) and comparing with the classical Heisenberg Hamiltonian, we extracted the interlayer exchange interaction $J_{\rm inter}$ depending on the layer separation ($d$). We found the oscillatory behavior of $J_{\rm inter}(d)$, well described by the RKKY model~\cite{Ruderman1954} (Fig. 2e). The systematic changes of $J_{\rm inter}(d)$ with Co doping, particularly between the nearest neighboring layers, determine the stability of the AFM phase. These results contrast to the case of insulating vdW antiferromagnets, in which the changes in the interlayer coupling from the FM to AFM types requires the stacking modifications~\cite{Sivadas2018, Jiang2019, Jang2019, Li2019, Song2019, Chen2019}, and manifest the important role of conduction electrons to control the spin configurations of (Fe$_{1-x}$Co$_x$)$_4$GeTe$_2$.

Conduction electrons are also important to probe the spin state of (Fe,Co)$_4$GeTe$_2$. With magnetic fields along the $c$ axis, normal to the preferred plane of the spin alignment, the moment at each layer of (Fe,Co)$_4$GeTe$_2$ gradually rotates until its mean direction is parallel to the field at $H^c_{\rm sat}$. This out-of-plane spin canting can be monitored by the anomalous Hall effect (AHE) due to sizable spin-orbit coupling in (Fe,Co)$_4$GeTe$_2$. As shown in Fig. 3a, the Hall resistivity $\rho_{yx}$ is dominated by the anomalous contribution $\rho^A_{yx}$ \textit{i.e.} $\rho_{yx}$  $\approx$ $\rho^A_{yx}$, and thus the transverse conductivity $\sigma_{yx}$ = $\rho_{yx}$/($\rho_{xx}^2$+$\rho_{yx}^2$) with variation of magnetic field is nicely scaled with $M(H)$. Moreover the scaling factor $S_{\rm H}$ = $\sigma_{yx}/M$ is found to be almost independent of temperature (Supplementary Fig. S3), and thus $\sigma_{yx}(H,T)$ can quantitatively measure the out-of-plane component of net magnetization $M(H,T)$. For example, the magnetic susceptibility, estimated from $\chi^c(T)$ $\propto$ $\sigma_{yx}(H)$/$H$, allows for experimentally determining $T_N$ in nanoflakes (Fig. 4).

The orientation of the staggered magnetization in the plane, \textit{i.e.}, the Neel vector in the plane can also be effectively tracked by the electrical conductivity. Figure 3b shows the field dependent magnetoresistance $\Delta \rho(H)/\rho(0)$ of (Fe,Co)$_4$GeTe$_2$ crystal under different field orientations in the plane, $H\parallel I$ and $H\perp I$ aginst a current $I \parallel a$ at various temperatures. Two characteristic fields, $H^{ab}_{\rm SF}$ and $H^{ab}_{\rm sat}$, are identified from clear kinks in both $\Delta \rho(H)/\rho(0)$ curves. The low field $H^{ab}_{\rm SF}$ corresponds to the spin-flop transition, at which the Neel vectors of all domains are fully aligned perpendicular to the in-plane field~\cite{Wang2019_2, Marti2014, Bodnar2018, Wadley2016, Galceran2016, Kriegner2016}. The Neel vector $L$ is rotated by 90$^{\circ}$, from $L\perp I$ to $L\parallel I$, by switching the in-plane magnetic field from $H \parallel I$ to $H \perp I$ (Fig. 3c). This leads to difference in magnetoresistance due to spin-orbit coupling, which is known as the anisotropic magnetoresistance (AMR) $\Delta \rho^L_{\rm AMR}$ = $\rho_{L \parallel I}$ $-$ $\rho_{L \perp I}$. In (Fe,Co)$_4$GeTe$_2$, $\Delta \rho^L_{\rm AMR}$ reaches $\sim$ 0.3\%, comparable with other AFM metals~\cite{Marti2014, Wadley2016, Galceran2016, Kriegner2016}.
With further increasing in-plane field, the moments are canted towards the field, until they are fully aligned at the saturation field $H^{ab}_{\rm sat}$. In this case, the AMR between the cases of $M \parallel I$ or $M \perp I$, $\Delta \rho^M_{\rm AMR}$ = $\rho_{M \parallel I}$ $-$ $\rho_{M \perp I}$ is also expected as found in ferromagnets~\cite{Sun2019, Takata2017}. In (Fe,Co)$_4$GeTe$_2$, $\Delta \rho^L_{\rm AMR}$ and $\Delta \rho^M_{\rm AMR}$ are comparable in size and opposite in sign, inducing the sign cross of $\Delta \rho_{\rm AMR}$ as summarized in Fig. 3d. Therefore, the AMR allows the electrical access to the orientation of the Neel vector or the saturated magnetization in the plane. Therefore combining AHE and AMR, we can effectively read out the spin state of (Fe,Co)$_4$GeTe$_2$.

Finally we discuss the thickness control of the magnetic state of (Fe,Co)$_4$GeTe$_2$.
Thanks to the weak vdW interlayer coupling, we obtained nanoflakes with thickness $d$ down to 7 layers, using mechanical exfoliation. The in-plane resistivity $\rho(T)$ as a function of temperature shows that
the overall resistivity increases with reducing thickness but retains the metallic behaviour ($\sim$ 10$^{-3}$ $\Omega$cm) (Fig. 4b). As in the bulk case (Fig. 3a), the transverse conductivity $\sigma_{yx}(H,T)$ gives the field and temperature dependent magnetization $M(H,T)$ and the susceptibility $\chi^c(T)$ of nanoflakes with $d$ = 7, 11 and 16 (Figs. 4c, 4d and Supplementary Fig. S8).
The resulting $\chi^c(T)$ exhibits a clear kink, indicating the AFM transition at $T_N$, which decreases gradually from the bulk value with lowering thickness. Additionally, in nanoflakes we found a broad hump developed around $T^*$ in $\chi^c(T)$, which shifts to low temperatures with reducing thickness (Fig. 4f). This nonmonotonous temperature dependence of $\chi^c(T)$ implies that spin moments are not fully frozen by dominant AFM interaction in nanoflakes, due to additional competing FM interaction, as discussed below.

The competing FM interaction stabilizes the long-range FM order with further reducing thickness to 7 layers (7L). The clear magnetic hysteresis of $\sigma_{yx}(H)$ with a coercive field $H_{\rm c}$ = 0.21 T, taken at $T$ = 2 K, evidences ferromagnetism in the 7L crystal (Fig. 4d).
The $H_{\rm c}$ gradually decreases with increasing temperature up to $T_c$ $\approx$ 25 K (Supplementary Fig. S8). The corresponding $\chi^c(T)$ is also distinct from those of the thicker crystals, showing no signature of AFM ordering but only a broad peak at $T^*$ $\sim$ 40 K.
The resulting phase diagram (Fig. 4f) reveals that thickness tuning offers another effective means to tune the interlayer FM and AFM interactions, even though the 2D limit is yet to be reached. It has been well established that vdW crystals expand along the $c$ axis by $\sim$ 0.3-0.7\%, ~\cite{Yoshida2014, Yoshida2017} when they are thinned to be tens of nanometer thick. Considering that the $c$-axis lattice constant shrinks by $\sim$ 0.6\% with Co doping of $x$ = 0.39, this thinning-induced swelling affects significantly the magnetic ground state, particularly near the FM-AFM phase boundary. Such a strong sensitivity of the magnetic phase with thickness variation is distinct from the insulating vdW magnets, whose magnetic phase is mostly maintained with thickness variation unless the stacking structure is changed.

Our observations unequivocally show that (Fe,Co)$_4$GeTe$_2$ is an intrinsic high-$T_N$ AFM multilayers.
Although its $T_N$ is still below room temperature, an order of magnitude enhancement of $T_N$, as compared to other metallic vdW antiferromagnets~\cite{Lei2020, Deng2020, Otrokov2019, Wu2019}, is achieved by tuning the vdW interlayer coupling between the strongly FM layers of Fe$_4$GeTe$_2$. This approach can be applied to other recently-discovered high-$T_c$ vdW ferromagnets~\cite{May2019, Stahl2018, Seo2020, Sun2019} to possibly realize the room temperature antiferromagnetism.
Furthermore, (Fe$_{1-x}$Co$_x$)$_4$GeTe$_2$ hosts at least four different spin states in terms of interlayer spin configurations (FM and AFM) and spin orientations (in-plane and out-of-plane) while keeping the same stacking structure (Fig. 2g).
The switching between these states, demonstrated using doping, magnetic field, and thickness control, can be more effectively achieved in vdW heterostructures
through \textit{e.g.} the exchange-spring effect~\cite{Park2011, Fina2014} with an adjacent FM vdW layer or the current-induced spin-orbit torque~\cite{Wadley2016, Bodnar2018} with large SOC materials, as done in metal spintronics.
The intimate coupling of the spin states to electrical conduction in (Fe,Co)$_4$GeTe$_2$ (Figs. 3a and 3b) ensures the electrical read out of the spin state. These strong controllability and readability on the spin states are manifestation of itinerant magnetism, highly distinct from insulating vdW magnets. We therefore envision that metallic vdW antiferromagets, including (Fe,Co)$_4$GeTe$_2$ in this work, will enrich the material candidates and the spin functionalities for all-vdW-material-based spintronics.

\subsection*{Methods}

\textbf{Single crystal growth and characterization.}
Single crystal were grown by a chemical vapor transport method using pre-synthesized polycrystalline sample and iodine as a transport agent. The obtained single crystals had platelike shape with a typical size of $\sim$ 1 $\times$ 1 $\times$ 0.04 mm$^3$. The high crytallinity of single crystals were confirmed by X-ray diffraction (Supplemenatary Fig. S1). From the energy dispersive spectroscopy measurements, we confirmed a systematic variation of Co doping $x$, in the presence of Te deficiency by $\sim$ 10\% and excess of the total (Fe, Co) content by $\sim$ 5\%, similar to pristine Fe$_4$GeTe$_2$~\cite{Seo2020}. Magnetization was measured under magnetic field along the $c$-axis or $ab$-plane using a superconducting quantum interference device (SQUID) magnetometer (MPMS, Quantum Design) and vibrating sample magnetometer (VSM) option of Physical Property Measurement System (PPMS-14T, Quantum Design). The in-plane resistivity and the Hall resistivity were measured in the standard six probe configuration using a Physical Property Measurement System (PPMS-9T, Quantum Design).

\textbf{Device fabrication.}
Using mechanical exfoliation in the inert argon atmosphere (H$_2$O $<$ 0.1 ppm, O$_2$ $<$ 0.1 ppm), we obtained nanoflakes of (Fe,Co)$_4$GeTe$_2$ on top of Si/SiO$_2$ substrate,
pre-treated by oxygen plasma (O$_2$ = 10 sccm, P $\sim$ 100 mTorr) for 5 minutes to remove adsorbates on the surface. The exfoliated crystal is then subsequently covered by a thin h-BN flake to prevent oxidation during device fabrication. Typically (Fe,Co)$_4$GeTe$_2$ flakes are of several tens of $\mu m^2$ in area and down to $\sim$ 7 nm ($\sim$ 7 layers) in thickness, estimated by the atomic force microscopy measurements (Supplementary Fig. S7). To make electrodes for transport measurements, we employed electron beam lithography technique, using poly(methyl methacrylate) (PMMA) positive resist layer, which was spin-coated and dried in vacuum at room temperature. After etching the patterned area of the covered h-BN flake with CF$_4$ plasma, Cr(10 nm)/Au(50 nm) are deposited on the exposed surface of the nanoflake.

\textbf{Scanning Transmission Electron Microscopy and Electron Energy Loss Spectroscopy.}
STEM samples were made along [110] projections which show the most distinguishable atomic structures. Samples were prepared with dual-beam focused ion beam(FIB) systems (Helios and Helios G3, FEI).
Different acceleration voltage conditions from 30 keV to 1 keV were used to make a thin sample with less damages.
Subsequent Ar ion milling process was performed with low energy (PIPS II, Gatan Inc., USA).
The atomic structure was observed using a STEM (JEM-ARM200F, JEOL, Japan) at 200 kV equipped with an 5th-order probe-corrector (ASCOR, CEOS GmbH, Germany).
The electron probe size was $\sim$0.8 ${\rm \AA}$, and the High Angle Annular Dark Field (HAADF) detector angle was fixed from 68 to 280 mrad.
The selected area diffraction pattern (SADP) was obtained using a TEM (JEM-2100F, JEOL, Japan) at 200 kV equipped with an spherical aberration corrector (CEOS GmbH, Germany).
The raw STEM images were compensated with 10 slices by SmartAlign and processed using a band-pass Difference filter with a local window to reduce a background noise (SmartAlign and Filters Pro, HREM research Inc., Japan).
For EELS analysis, we utilized another STEM (JEM-2100F, JEOL) with a spherical aberration corrector (CEOS GmbH) equipped with an EEL spectrometer (GIF Quantum ER, Gatan, USA). The used probe size was $\sim$1.0 ${\rm \AA}$ under 200 kV and the chemical analysis was performed by using a Spectrum Image via the STEM mode. The obtained spectral data from a SI were filtered to intensify the Fe-, Co-, and Te-edge signals by MSA (Multivariate Statistical Analysis, HREM research Inc., Japan).

\textbf{Scanning Tunneling Microscopy.}
The surface of (Fe,Co)$_4$GeTe$_2$ was characterized by  the scanning tunneling miscroscopy (STM) on
single crystal. The single crystal is first cleaved in high vacuum ($\sim$ 1 $\times$ 10$^{-8}$ torr) to obtain
the clean surface and then is transferred to ultra high vacuum ($\sim$ 1 $\times$ 10$^{-11}$ torr) for the STM
measurements at 77 K.

\textbf{First principles calculations.}
Electronic structures calculation were performed using a full-potential linearized augmented plane wave method, implemented in WIEN2K package ~\cite{Blaha2001}. Experimental lattice constants of (Fe,Co)$_4$GeTe$_2$ are used and exchange-correlation potential is chosen to the generalized gradient approximation of Perdew-Berke-Ernzerhof (PBE-GGA)~\cite{PBE}. Assuming that Co is doped at all Fe sites homogenously throughout the Fe$_4$GeTe$_2$ layer, the virtual crystal approximation is considered in the DFT calculation of (Fe,Co)$_4$GeTe$_2$.
\\

%===========================
%\pagebreak
\let\origdescription\description
\renewenvironment{description}{
\setlength{\leftmargini}{0em}
\origdescription
\setlength{\itemindent}{0em}
\setlength{\labelsep}{\textwidth}
}
{\endlist}

\begin{description}
\item[Acknowledgment] The authors thank H. W. Lee, K. Kim, M. H. Jo for fruitful discussion. We also thank H. G. Kim in Pohang Accelerator Laboratory (PAL) for the technical support. This work was supported by the Institute for Basic Science (IBS) through the Center for
Artificial Low Dimensional Electronic Systems (no. IBS-R014-D1), and by the National Research Foundation of Korea (NRF) through SRC (Grant No. 2018R1A5A6075964), the Max Planck-POSTECH Center for Complex Phase Materials (Grant No. 2016K1A4A4A01922028). S.Y.C. acknowledges the support of the Global Frontier Hybrid Interface Materials by the NRF (Grant No. 2013M3A6B1078872). K.W. and T.T. acknowledge support from the Elemental Strategy Initiative conducted by the MEXT, Japan ,Grant Number JPMXP0112101001, JSPS KAKENHI Grant Numbers JP20H00354 and the CREST(JPMJCR15F3), JST. J.J acknowlege the support from the NRF (Grand No. NRF-2019R1A2C1089017 and NRF-2018K2A9A1A06069211).

\item[Author Contributions] J.S.K., E.S.A. and J.S conceived the experiments. J.S synthesized the bulk crystals. E.S.A. carried out device fabrication and measurements on nanoflakes. J.S., M.H.C. and Y.J.J performed the transport property measurements on bulk crystals. T.P. and J.H.S. performed the electronic structure calculations and the analysis. S.-Y.H., G.-Y.K., K.S. and S.-Y.C. carried out structural and chemical identification using scanning transmission electron microscopy and electron energy loss spectroscopy. E.O. and H.W.Y. performed surface characterization using scanning tunneling microscopy measurements. K.W and T.T provided boron nitride crystals. J.S., E.S.A., T.P., S.-Y.C., J.H.S. and J.S.K. co-wrote the manuscript. All authors discussed the results and commented on the paper.

\item[Competing financial interests] The authors declare no competing
financial interests.

\item[Additional information] Correspondence and requests for materials should be addressed to S. -Y. Choi~(email: youngchoi@postech.ac.kr), J. H. Shim~(email: jhshim@postech.ac.kr) and J. S. Kim~(email: js.kim@postech.ac.kr).
\end{description}

\pagebreak

%%%%%%%%%%%%%%%%%%%%%%%%%%%%%%%%%%%%%%%%%%%%%%%%%%%%%%%%%%%%%%%%%%%
\begin{figure*}[t]
\centering
\includegraphics [width=13 cm, bb= 20 300 520 750]{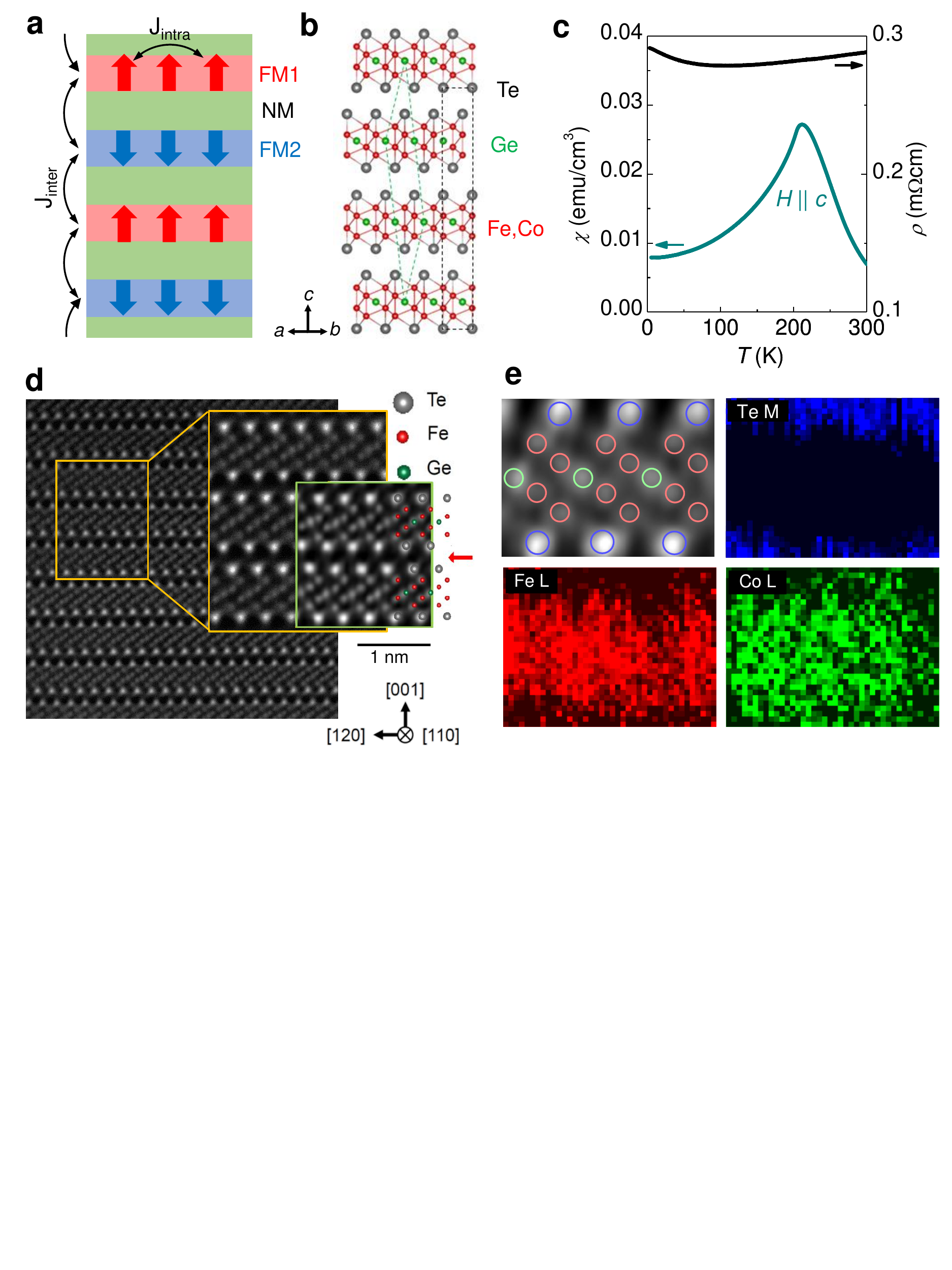}
\caption{
{\bf Crystal structure and antiferromagnetism of (Fe,Co)$_4$GeTe$_2$.}
{\bf a,b}, Schematic illustration of synthetic antiferromagnetic (AFM) multilayers and crystal structure of (Fe,Co)$_4$GeTe$_2$. In both cases, ferromagnetic (FM) layers with intralayer exchange coupling ($J_{\rm intra}$) are stacked alternately and coupled by interlayer coupling ($J_{\rm inter}$).
{\bf c}, Temperature dependent magnetic susceptibility $\chi(T)$ and $\rho (T)$ of (Fe$_{1-x}$Co$_x$)$_4$GeTe$_2$ for $x$ = 0.33. A clear cusp in $\chi(T)$ reveals the AFM transition at $T_N$ $\approx$ 210 K.
{\bf d}, HAADF Scanning transmission electron microscopy (STEM) image of (Fe,Co)$_4$GeTe$_2$ (the first two ones) and Fe$_4$GeTe$_2$ (the last one) crystals along [110]. The STEM image with lower magnification shows all layers are regularly arranged with clear vdW gap (red arrow) and without stacking faults throughout whole areas. The comparison of STEM images with higher magnification of (Fe,Co)$_4$GeTe$_2$ (yellow box) and Fe$_4$GeTe$_2$ (green box) show the almost identical atomic structure.
{\bf e}, EELS intensity distributions of Te L, Fe L, and Co L edges within the monolayer of (Fe,Co)$_4$GeTe$_2$, indicating the possibilities of finding the corresponding atoms. Large intensity of Te L edge is found at the top and bottom of a monolayer as expected, whereas the intensity for Fe and Co atoms are uniformly distributed between Te layers. This infers that Co dopants are successfully substituting Fe atoms as solid solution.
} \label{fig1}
\end{figure*}
%%%%%%%%%%%%%%%%%%%%%%%%%%%%%%%%%%%%%%%%%%%%%%%%%%%%%%%%%%%%%%%%%%%

%%%%%%%%%%%%%%%%%%%%%%%%%%%%%%%%%%%%%%%%%%%%%%%%%%%%%%%%%%%%%%%%%%%
\begin{figure*}[t]
\centering
\includegraphics[width=16cm, bb=10 450 540 720]{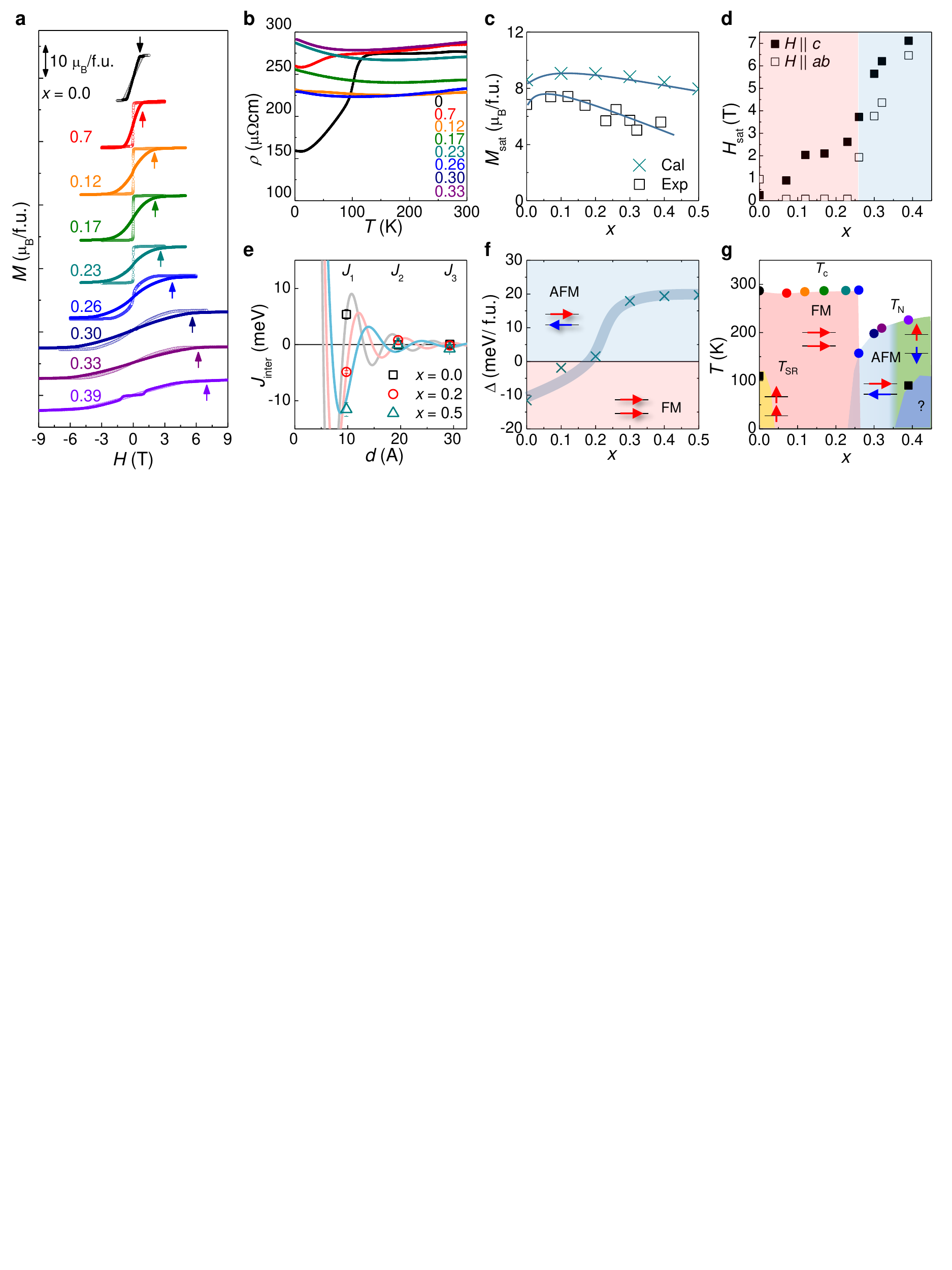}
\caption{
{\bf Doping dependent magnetic phase diagram of (Fe$_{1-x}$Co$_x$)$_4$GeTe$_2$.}
{\bf a}, Magnetization $M(H)$ as a function of magnetic field for (Fe$_{1-x}$Co$_x$)$_4$GeTe$_2$ (0 $\leq$ x $\leq$ 0.39) single crystals, for different field orientations, $H\parallel c$ (solid) and $H\parallel ab$ (open). All the $M(H)$ curves were taken at $T$ = 10 K, except those for $x$ = 0.39 taken at $T$ = 100 K.
{\bf b}, Temperature dependent in-plane resistivity $\rho(T)$ showing a metallic behavior.
{\bf c,d}, The saturation magnetization $M_{\rm sat}$ ({\bf c}) and fields $H_{\rm sat}$ ({\bf d}) for $H\parallel c$ (solid) and $H\parallel ab$ (open) as a function of Co doping $x$. The calculated $M_{\rm sat}$ is also shown in {\bf c} for comparison.
{\bf e}, Calculated interlayer exchange interaction $J_{\rm inter}$ with different layer distance ($d$) for $x$ = 0, 0.2, and 0.5. The oscillatory dependence of $J_{\rm inter}(d)$, captured by the RKKY model, is systematically changed with Co doping.
{\bf f}, Doping dependent total energy difference $\Delta$ between the FM and the A-type AFM phases. The FM-to-AFM transition occurs at the critical doping level $x_c$ $\sim$ 0.2, in good agreement with experiments.
{\bf g}, Phase diagram of spin states for (Fe$_{1-x}$Co$_x$)$_4$GeTe$_2$ with Curie ($T_C$) and Neel ($T_N$) temperatures.
Four different spin states in terms of interlayer spin configurations (FM and AFM) and spin orientations (easy-axis and easy-plane) are stabilized depending on doping level and temperature. The unknown magnetic phase is found for $x$ = 0.39 below $T$ = 90 K. The schematic illustration of spin configurations are shown in the inset.
} \label{fig2}
\end{figure*}
%%%%%%%%%%%%%%%%%%%%%%%%%%%%%%%%%%%%%%%%%%%%%%%%%%%%%%%%%%%%%%%%%%%

%%%%%%%%%%%%%%%%%%%%%%%%%%%%%%%%%%%%%%%%%%%%%%%%%%%%%%%%%%%%%%%%%%%
\begin{figure*}[t]
\centering
\includegraphics[width=13 cm, bb=0 350 530 750]{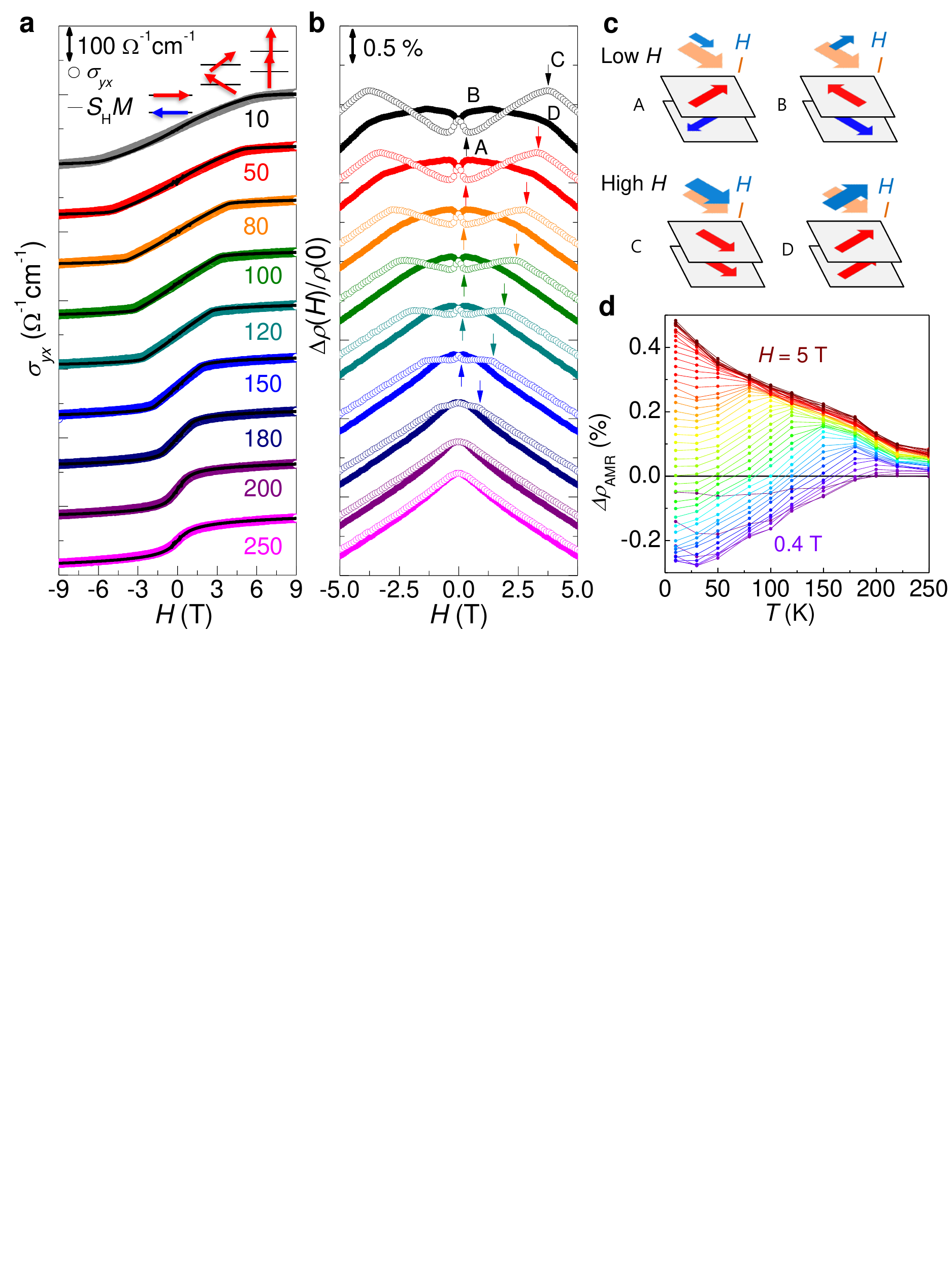}
\caption{
{\bf Electrical detection of the spin states.}
{\bf a}, The transverse conductivity $\sigma_{yx}(H)$ as a function of magnetic fields along the $c$ axis, taken at various temperatures. The $\sigma_{yx}(H)$ data are nicely reproduced by the field dependent magnetization $M(H)$ (black solid line) with a scaling factor $S_{\rm{H}}$ $\approx$ 0.3 V$^{-1}$, following the linear relation of $\sigma_{yx}(H)$ = $S_{{\rm H}}M(H)$.
{\bf b}, Magnetoresistance $\Delta\rho(H)/\rho(0)$ under in-plane magnetic fields $H$, parallel (open) or perpendicular (solid) to the current $I$ along the $a$ axis. The low field spin-flop transition field $H^{ab}_{\rm SF}$ and the high field saturation field $H^{ab}_{\rm sat}$ are indicated by the arrows.
{\bf c}, Spin configurations with different relative orientations of the magnetic field $H$ and the current $I$. At low $H$, the antiferromagnetically coupled spins are aligned perpendicular to $H$, either $H \parallel I$ (A) or $H \perp I$ (B). At high $H$, the saturated spins are aligned parallel to $H$, either $H \parallel I$ (C) or $H \perp I$ (D).
{\bf d}, Anisotropic magnetoresistance $\Delta \rho_{\rm AMR}$ as a function of temperature and in-plane magnetic field. The low-field and high-field AMR, determined by the relative orientation of Neel vector and the saturated magnetization against the current direction, respectively, which results in a sign change.
} \label{fig3}
\end{figure*}
%%%%%%%%%%%%%%%%%%%%%%%%%%%%%%%%%%%%%%%%%%%%%%%%%%%%%%%%%%%%%%%%%%%

%%%%%%%%%%%%%%%%%%%%%%%%%%%%%%%%%%%%%%%%%%%%%%%%%%%%%%%%%%%%%%%%%%%
\begin{figure*}[t]
\centering
\includegraphics[width=16 cm, bb=0 350 550 700]{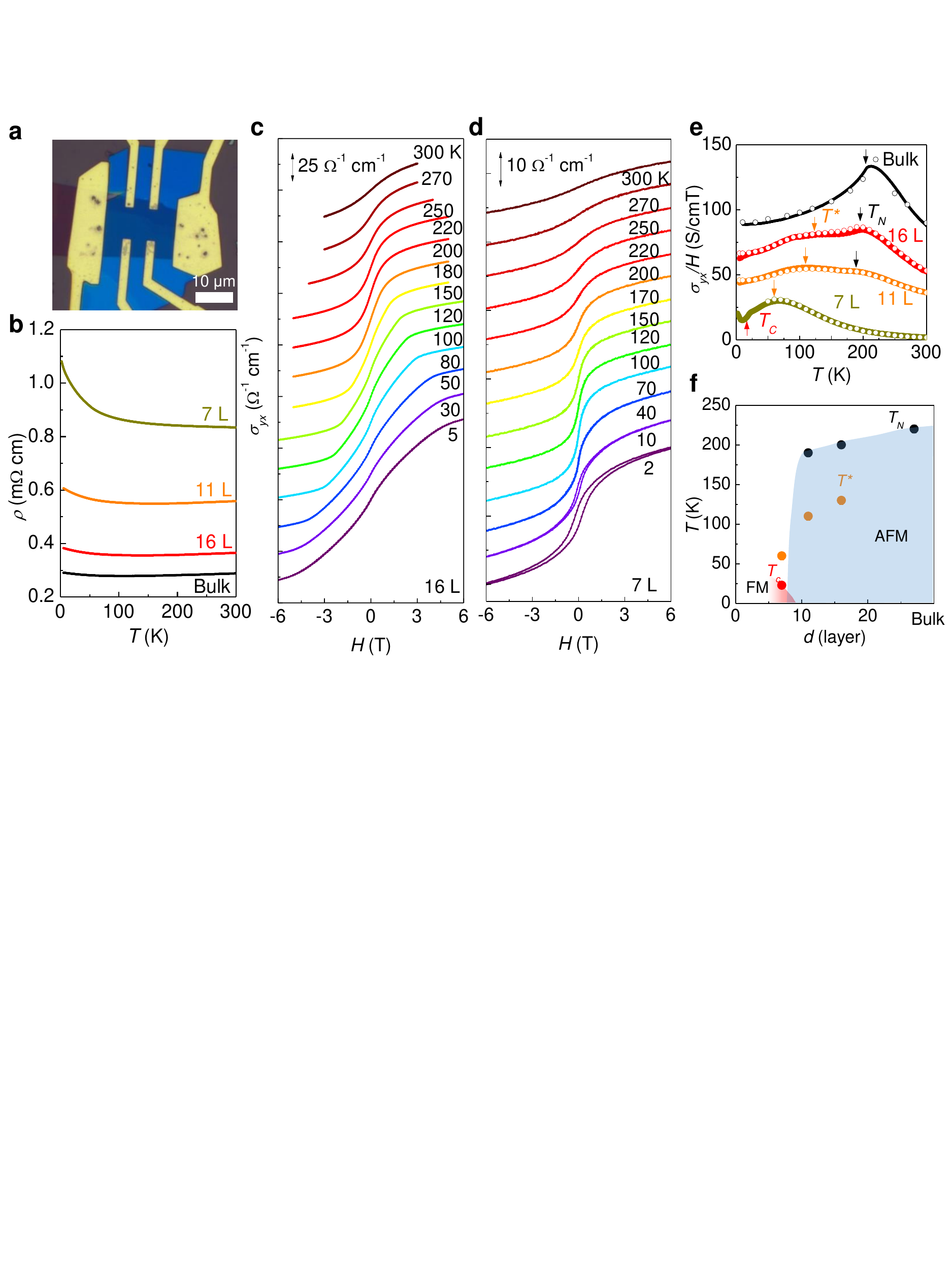}
\caption{
{\bf Thickness-dependent magnetic phase diagram.}
{\bf a}, The optical image of BN-covered 7-layer-thick (7 L) (Fe,Co)$_4$GeTe$_2$ crystal with the scale bar of 10 $\mu$m.
{\bf b}, Temperature dependent in-plane resistivity $\rho(T)$ for bulk and nanoflakes with various thickness, indicated by number of the layers.
{\bf c,d}, Magnetic field dependent transverse conductivity $\sigma_{yx}(H)$ at various temperatures for $H \parallel c$ in 16L ({\bf c}) and 7L nanoflakes ({\bf d}). The $\sigma_{yx}(H)$ curves at low temperatures resembles typical field-dependent magnetization for the AFM and FM phases in {\bf c} and {\bf d}, respectively.
{\bf e}, Temperature dependence of the magnetic susceptibility $\chi^c(T)$ for nanoflakes and bulk, estimated from the low-field slope of $\sigma_{yx}(H)$ at each temperature (open circle) or from the difference between $\sigma_{yx}(T)$ curves taken at $H$ = $\pm$ 0.1 T for $H \parallel c$. Magnetic transition temperatures $T_N$ and $T_C$ are indicated by the arrows, together with the characteristic temperature $T^*$, determined by a broad hump in $\chi^c(T)$.
{\bf f}, Thickness dependent magnetic phase diagram with characteristic temperatures of $T_N$, $T^*$, and $T_c$.
} \label{fig4}
\end{figure*}
%%%%%%%%%%%%%%%%%%%%%%%%%%%%%%%%%%%%%%%%%%%%%%%%%%%%%%%%%%%%%%%%%%%

\end{document}